\documentclass[pra,twocolumn]{revtex4}
\usepackage{graphicx}
\begin{document}

\bibliographystyle{unsrt}

\title{Demonstration of Feed-Forward Control for Linear Optics Quantum Computation}
\author{T.B. Pittman, B.C. Jacobs, and J.D. Franson}
\affiliation{The Johns Hopkins University,
Applied Physics Laboratory, Laurel, MD 20723}

\date{\today}

\begin{abstract}
One of the main requirements in linear optics quantum computing is the ability to perform single-qubit operations that are controlled by classical information fed forward from the output of single photon detectors. These operations correspond to pre-determined combinations of phase corrections and bit-flips that are applied to the post-selected output modes of non-deterministic quantum logic devices.  Corrections of this kind are required in order to obtain the correct logical output for certain detection events, and their use can increase the overall success probability of the devices.  In this paper, we report on the experimental demonstration of the use of this type of feed-forward system to increase the probability of success of a simple non-deterministic quantum logic operation from approximately $\frac{1}{4}$ to $\frac{1}{2}$. This logic operation involves the use of one target qubit and one ancilla qubit which, in this experiment, are derived from a parametric down-conversion photon pair.  Classical information describing the detection of the ancilla photon is fed-forward in real-time and used to alter the quantum state of the output photon.  A fiber optic delay line is used to store the output photon until a polarization-dependent phase shift can be applied using a high speed Pockels cell.
\end{abstract}

\maketitle

\section{Introduction}
\label{sec:intro}

In a recent paper, Knill, LaFlamme, and Milburn (KLM) showed that efficient quantum computation is possible using only linear optical elements, ancilla photons, and post-selection based on the outcome of single-photon detectors \cite{knill01a}.  Roughly speaking, measurements made on the ancilla photons will project out the desired logical output state provided that certain measurement results were obtained, which only occurs for some fraction of the events.  Additional events can be accepted as well if single-bit corrections (phase shifts and bit-flips) are applied to the output qubits based on the results of the ancilla measurements, which increases the overall probability of success.  Here we report the experimental demonstration of feed-forward control operations of this kind that were used to increase the probability of success of a simple quantum logic operation from $\frac{1}{4}$ to $\frac{1}{2}$.

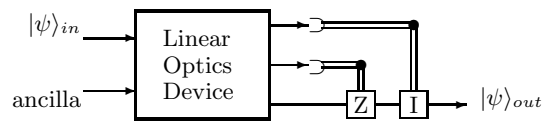
\begin{figure}[b]
\begin{center}
\begin{picture}(200,75)
\thicklines
\put(50,10){\framebox(50,40)}
\thinlines
\put(60,37){Linear}
\put(60,27){Optics}
\put(60,17){Device}
\put(30,40){\vector(1,0){20}}
\put(30,20){\vector(1,0){20}}
\put(3,14){ancilla}
\put(9,43){$|\psi\rangle_{in}$}
\put(100,45){\vector(1,0){15}}
\put(115,43){$\supset$}
\put(100,30){\vector(1,0){15}}
\put(115,27){$\supset$}
\put(100,15){\line(1,0){30}}
\put(130,10){\framebox(10,10){Z}}
\put(140,15){\line(1,0){10}}
\put(150,10){\framebox(10,10){I}}
\put(160,15){\vector(1,0){15}}
\put(180,15){$|\psi\rangle_{out}$}
\put(120,46){\line(1,0){35}}
\put(120,44){\line(1,0){35}}
\put(154,20){\line(0,1){25}}
\put(156,20){\line(0,1){25}}
\put(155,45){\circle*{4}}
\put(120,31){\line(1,0){15}}
\put(120,29){\line(1,0){15}}
\put(134,20){\line(0,1){10}}
\put(136,20){\line(0,1){10}}
\put(135,30){\circle*{4}}
\end{picture}
\end{center}
\vspace*{-.25in}
\caption{A simple example of a non-deterministic quantum logic operation demonstrating the use of feed-forward control. Here the logic operation uses a single ancilla photon and post-selection to perform a unitary transformation on an arbitrary input state of another single photon $|\psi\rangle_{in}$.  Classical information describing the outcome of two single-photon detectors is fed forward (along double wires) to control units which perform, for example, the single-qubit operations $I$ or $Z$ on the accepted output.  The notation follows that of reference \protect\cite{chuangneilsen}.}
\label{fig:basicidea} 
\end{figure}

An example of a simple feed-forward control process is illustrated in Figure \ref{fig:basicidea}.  Here the results of measurements made on a single ancilla photon are used to apply one of two possible single-qubit transformations to the output.  It can be seen that the feed-forward control process required for probabilistic quantum logic gates is similar to the unitary transformations that would be required for a complete implementation of conventional quantum teleportation \cite{bennett93}, given the results of a Bell-state measurement.  Feed-forward control would also be needed for a variety of other quantum optics proposals (see, for example, \cite{lutkenhaus99,kok00}).

The original proposal \cite{knill01a,knill01b} for a non-deterministic gate was based on an interferometer arrangement whose stability was subsequently improved by Ralph {\em et.al.} \cite{ralph01a}.  The experiments reported here are based on the use of polarization-encoded qubits and polarizing beamsplitters \cite{pan01a,koashi01}, which we previously used to demonstrate several non-deterministic quantum logic devices \cite{pittman01a,pittman01b}.  A variety of other types of probabilistic quantum logic operations have also been described \cite{koashi01,rudolph01,zou01}, and schemes which illustrate the basic properties of non-deterministic logic gates in the coincidence basis have also been proposed  \cite{ralph01b,hofmann01}.

The non-deterministic logic operation chosen for this particular demonstration was a probabilistic quantum parity check \cite{pittman01a,pittman01b}, but the techniques and results presented here are expected to apply to other non-deterministic logic devices as well.  The quantum parity check was chosen because of its 
relatively simple structure, which involves an input of only one target qubit and one ancilla qubit in analogy with the example shown in Figure 1.  In our experiment, these two qubits are derived from a parametric down-conversion pair, and the detection of the ancilla photon by one of two detectors determines which single-qubit operation needs to be applied to the output photon. The output photon was stored for roughly 100 ns using a fiber optic delay line while the classical detection signal was amplified and fed forward to a Pockels cell that was used to apply a state-dependent phase shift.

The remainder of the paper is outlined as follows: in Section \ref{sec:theory} we review the  goals and theory of operation of the quantum parity check, highlighting the need for feed-forward controlled single-qubit operations.  In Section \ref{sec:experiment} we describe the details of the experiment and present the results.  In Section \ref{sec:summary} we summarize and discuss the need for feed-forward control for more general applications in a linear optics quantum computing protocol.

\section{Probabilistic Quantum Parity Check}
\label{sec:theory}

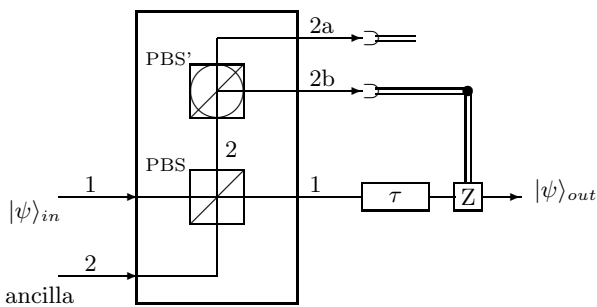
\begin{figure}[b]
\begin{center}
\begin{picture}(230,130)
\thicklines
\put(50,10){\framebox(60,110)}
\thinlines
\put(70,40){\framebox(20,20)}
\put(70,40){\line(1,1){20}}
\put(70,80){\framebox(20,20)}
\put(70,80){\line(1,1){20}}
\put(80,90){\circle{20}}
\put(53,100){\scriptsize PBS'}
\put(53,60){\scriptsize PBS}
\put(50,50){\line(1,0){60}}
\put(50,20){\line(1,0){30}}
\put(80,20){\line(0,1){90}}
\put(80,110){\line(1,0){30}}
\put(80,90){\line(1,0){30}}
\put(83,65){2}
\put(20,50){\vector(1,0){30}}
\put(20,20){\vector(1,0){30}}
\put(0,10){ancilla}
\put(1,42){$|\psi\rangle_{in}$}
\put(30,52){1}
\put(30,22){2}
\put(110,110){\vector(1,0){25}}
\put(115,112){2a}
\put(135,108){$\supset$}
\put(110,90){\vector(1,0){25}}
\put(115,92){2b}
\put(135,88){$\supset$}
\put(110,50){\line(1,0){25}}
\put(115,52){1}
\put(135,45){\framebox(25,10){$\tau$}}
\put(160,50){\line(1,0){10}}
\put(170,45){\framebox(10,10){Z}}
\put(180,50){\vector(1,0){15}}
\put(200,50){$|\psi\rangle_{out}$}
\put(140,111){\line(1,0){15}}
\put(140,109){\line(1,0){15}}
\put(174,55){\line(0,1){35}}
\put(176,55){\line(0,1){35}}
\put(140,91){\line(1,0){35}}
\put(140,89){\line(1,0){35}}
\put(175,90){\circle*{4}}
\end{picture}
\end{center}
\vspace*{-.25in}
\caption{Implementation of a probabilistic quantum parity check using a polarizing beamsplitter and a polarization-sensitive detection package. The post-selection process involves accepting the output if exactly one photon is registered by one of the detectors in modes $2a$ or $2b$.  If the photon is found in mode $2a$ the desired logical output is obtained, while a state-dependent $\pi$ phase shift ($Z$-operation) is required if the photon is found in mode $2b$. From an experimental point of view, the output photon must be delayed by some time interval $\tau$ while the classical information is processed and used to apply the single-qubit $Z$-operation. }
\label{fig:qpc} 
\end{figure}

The implementation of a probabilistic quantum parity check is shown in Figure \ref{fig:qpc}. The basic theory of its operation has been presented elsewhere \cite{pittman01a}, but will be briefly reviewed here for self-consistency and to emphasize the use of classically controlled single-qubit operations.

The quantum parity check utilizes polarization-encoded qubits \cite{bouwmeester2000a}, where the logical values $0$ and $1$ are represented by the horizontal and vertical polarization states of a single photon. When the qubits have values of either $0$ or $1$, the goal of the quantum parity check is to transfer the value of the qubit in mode $1$ to the output, provided that its value is the same as that of an ancilla qubit in mode $2$ (even parity). The device fails and produces no output if these two qubits have opposite values (odd parity). Basis-state parity checks of this kind have been found to be extremely useful in a variety of quantum information processing applications (see, for example, \cite{pan98,pan01a,pan01b,bouwmeester01,yamamoto01,zhao01}).  In particular, we have shown that the probabilistic quantum parity check shown in Figure \ref{fig:qpc} can form the basis of a quantum encoding operation and a non-deterministic controlled-NOT gate \cite{pittman01a}.

The operation of the parity check is more subtle when the two input states are superpositions of $0$ and $1$.  In particular, the case in which the ancilla photon is prepared in an equal superposition of the computational basis states, 
$|\psi\rangle_{anc} =\frac{1}{\sqrt{2}}(|H_{2}\rangle+|V_{2}\rangle)$ 
is of special interest.  In this case, the quantum parity check is able to coherently transfer any arbitrary superposition state of the input qubit,  
$|\psi\rangle_{in}\equiv \alpha|H_{1}\rangle+\beta|V_{1}\rangle$,
 into the output.  This aspect of the quantum parity check device will be used for the demonstration of our feed-forward control system.

As shown in Figure \ref{fig:qpc}, the operation of the non-deterministic quantum parity check consists of mixing the target photon and the ancilla photon in a polarizing beamsplitter ($PBS$), and the post-selection process involves accepting the output if and only if exactly one photon is received by one of the two single-photon detectors in modes $2a$ and $2b$.
Note that the $PBS$ would ordinarily transmit only the horizontal component 
$\alpha|H_{1}\rangle$ of the input state into the output (the vertical component is reflected into mode $2$).  However, the use of the prepared ancilla photon and the post-selection process essentially render the $PBS$ ``transparent'' to the entire target state.  This nonclassical action can be understood by considering the transformation of the total state,
$|\psi_{T}\rangle \equiv |\psi\rangle_{in} \otimes |\psi\rangle_{anc}$,
by the $PBS$:

\begin{equation}
|\psi_{T}\rangle \rightarrow 
\frac{\alpha}{\sqrt{2}} |H_{1}\rangle|H_{2}\rangle +
\frac{\beta}{\sqrt{2}} |V_{1}\rangle|V_{2}\rangle +
\frac{1}{\sqrt{2}}|\psi_{\perp}\rangle
\label{eq:qpc1}
\end{equation}

\noindent
where 
$|\psi_{\perp}\rangle \equiv \alpha|H_{1}\rangle|V_{1}\rangle +
 \beta|H_{2}\rangle|V_{2}\rangle$ includes amplitudes which cannot satisfy the post-selection criterion and are therefore rejected.

From equation (\ref{eq:qpc1}) we see that direct measurement of the polarization of the photon in mode $2$ would determine the value of the input qubit and destroy the coherence of any subsequent operations.  For this reason, an additional polarizing beamsplitter ($PBS'$) is placed in mode $2$ and oriented in a basis that is rotated by $45^{o}$ from the horizontal-vertical basis. In this way the detection of the photon in modes $2a$ or $2b$ provides no information regarding its origin.

Expanding the relevant terms of equation (\ref{eq:qpc1}) in the $45^{o}$ basis of the detector modes shows that:

\begin{equation}
|\psi_{T}\rangle \rightarrow 
\frac{1}{2}\left[|D_{2a}\rangle(\alpha|H_{1}\rangle+\beta|V_{1}\rangle)
+|D_{2b}\rangle(-\alpha|H_{1}\rangle+\beta|V_{1}\rangle)\right]
\label{eq:qpc2}
\end{equation}

\noindent
where, for example, $|D_{2b}\rangle$ represents a single photon in detector mode $2b$. This state is unnormalized due to the rejection of $|\psi_{\perp}\rangle$, which is responsible for the probabilistic nature of the device. 

Equation (\ref{eq:qpc2}) illustrates the need for feed-forward and classically controlled single-qubit operations. Note that if the ancilla photon is registered by the detector in mode $2a$ then the output is projected into the desired state, whereas if the photon is found in mode $2b$ the projected output state requires a  $\pi$ phase shift on the $|H_{1}\rangle$ component relative to the $|V_{1}\rangle$ component:

\begin{eqnarray}
&D_{2a} \Longrightarrow |\psi\rangle_{out}=\alpha|H_{1}\rangle+\beta|V_{1}\rangle
\nonumber \\
&D_{2b} \Longrightarrow |\psi\rangle_{out}=-\alpha|H_{1}\rangle+\beta|V_{1}\rangle
\label{eq:qpc3}
\end{eqnarray}

\noindent 
This state-dependent $\pi$ phase shift is equivalent to applying the Pauli $\hat{\sigma}_{z}$ spin operator and is often referred to as a single-qubit $Z$ operation \cite{chuangneilsen}.

Each of the outcomes described in equation (\ref{eq:qpc3}) is equally likely and occurs with a probability of $\frac{1}{4}$.  Therefore, if the device is passively run such that only detections in mode $2a$ are accepted, the success probability of the logic operation is $\frac{1}{4}$. This aspect of the quantum parity check has been experimentally verified \cite{pittman01b}.  However, if the other detection outcome is also accepted, and the classically controlled single-qubit $Z$-operation is successfully implemented, the overall success probability of the gate is increased to $\frac{1}{2}$. As will be described in the next section, an  experimental implementation of this procedure is the main result of this paper.

\section{Experiment and Results}
\label{sec:experiment}

\subsection{Experimental Design}
\label{sec:schematic}

\begin{figure}[t]
\begin{center}
\begin{picture}(230,170)
\thicklines
\put(60,20){\framebox(20,20)}
\put(60,20){\line(1,1){20}}
\thinlines
\put(49,42){\scriptsize PBS}
\put(60,30){\line(1,0){20}}
\put(70,20){\line(0,1){20}}
\put(0,15){\scriptsize photon}
\put(3,9){\scriptsize pair}
\put(20,30){\vector(1,0){15}}
\put(35,25){\framebox(4,10)}
\put(34,39){\scriptsize $\frac{\lambda}{2}$}
\put(39,30){\line(1,0){21}}
\put(48,23){\scriptsize 1}
\put(20,0){\vector(1,0){15}}
\put(35,-5){\framebox(4,10)}
\put(34,-14){\scriptsize $\frac{\lambda}{2}$}
\put(39,0){\line(1,0){31}}
\put(70,0){\line(0,1){20}}
\put(73,6){\scriptsize 2}
\put(80,30){\vector(1,0){10}}
\put(91,28){$>$}
\put(92,20){\scriptsize fc}
\put(95,24){\oval(16,12)[tr]}
\put(111,24){\oval(16,12)[bl]}
\put(111,18){\line(1,0){79}}
\put(125,28){\circle{18}}
\put(127,28){\circle{18}}
\put(129,28){\circle{18}}
\put(120,10){\scriptsize fiber}
\put(110,4){\scriptsize delay line}
\put(150,18){\framebox(24,8)}
\put(154,22){\circle{8}}
\put(162,22){\circle{8}}
\put(170,22){\circle{8}}
\put(157,10){\scriptsize fpc}
\put(190,26){\oval(20,16)[br]}
\put(197,26){$\vee$}
\put(205,28){\scriptsize fc}
\put(200,33){\vector(0,1){17}}
\put(193,50){\framebox(14,20){\scriptsize PC}}
\put(200,70){\line(0,1){10}}
\put(192,80){\framebox(16,4)}
\put(192,80){\line(4,1){16}}
\put(211,80){\scriptsize $\theta_{1}$}
\put(200,84){\line(0,1){10}}
\put(196,94){\framebox(8,2)}
\put(207,93){$f$}
\put(200,96){\vector(0,1){10}}
\put(197,108){$\cap$}
\put(205,110){\scriptsize $D_{1}$}
\put(70,40){\line(0,1){14}}
\put(66,54){\framebox(8,2)}
\put(77,53){$f$}
\put(70,56){\line(0,1){9}}
\put(65,65){\framebox(10,4)}
\put(57,65){\scriptsize $\frac{\lambda}{2}$}
\thicklines
\put(60,75){\framebox(20,20)}
\put(60,75){\line(1,1){20}}
\thinlines
\put(40,80){\scriptsize PBS'}
\put(70,69){\vector(0,1){37}}
\put(67,108){$\cap$}
\put(52,110){\scriptsize $D_{2a}$}
\put(70,85){\line(1,0){20}}
\put(90,85){\vector(0,1){21}}
\put(87,108){$\cap$}
\put(94,110){\scriptsize $D_{2b}$}
\put(69,113){\line(0,1){22}}
\put(71,113){\line(0,1){20}}
\put(69,135){\line(1,0){64}}
\put(71,133){\line(1,0){62}}
\put(89,113){\line(0,1){12}}
\put(91,113){\line(0,1){10}}
\put(89,125){\line(1,0){44}}
\put(91,123){\line(1,0){42}}
\put(115,110){\framebox(40,50)}
\put(117,151){\scriptsize TTL logic}
\put(122,145){\scriptsize board}
\put(133,118){\framebox(14,20){\scriptsize OR}}
\put(123,124){\circle*{4}}
\put(122,59){\line(0,1){65}}
\put(124,61){\line(0,1){63}}
\put(124,61){\line(1,0){21}}
\put(122,59){\line(1,0){23}}
\put(145,50){\framebox(36,20){\scriptsize PC driver}}
\put(181,59){\line(1,0){12}}
\put(181,61){\line(1,0){12}}
\put(167,140){\framebox(45,20)}
\put(170,151){\scriptsize coincidence}
\put(170,145){\scriptsize logic circuit}
\put(199,113){\line(0,1){27}}
\put(201,113){\line(0,1){27}}
\put(147,130){\line(1,0){32}}
\put(147,128){\line(1,0){34}}
\put(179,130){\line(0,1){10}}
\put(181,128){\line(0,1){12}}
\end{picture}
\end{center}
\caption{A schematic of the experiment designed to demonstrate the use of feed-forward control in a basic non-deterministic quantum logic device. The target and ancilla photons derived from a parametric down-conversion pair are sent into a quantum parity check device in analogy with the schematic of Figure \protect\ref{fig:qpc}.  $PBS$ and $PBS'$ are polarizing beamsplitters, while the $\frac{\lambda}{2}$ are half-waveplates used for state preparation. $D_{1}$, $D_{2a}$, and $D_{2b}$ are single-photon detectors, and $f$ represents 10 nm bandwidth filters centered at the wavelength of the down-converted photons.  The fc units are fiber couplers and fpc is a fiber polarization controller. The logical output state is verified by polarization analyzer $\theta_{1}$, and the classically controlled $Z$-operation described in Section  \protect\ref{sec:theory} is implemented with a Pockels cell (PC).}
\label{fig:experiment} 
\end{figure}
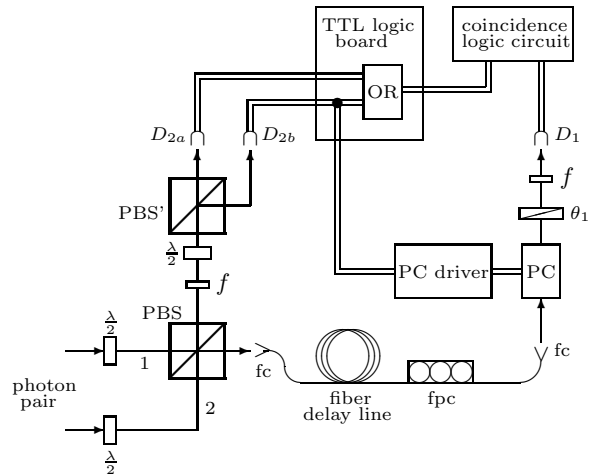

A schematic of the experimental set up used to demonstrate the use of feed-forward and classically controlled single-qubit operations in the non-deterministic quantum parity check is shown in Figure \ref{fig:experiment}. As mentioned earlier, the target and ancilla photons are derived from a conventional parametric down-conversion photon pair source \cite{shih94} which is not shown in Figure \ref{fig:experiment}, but is described in detail in reference \cite{pittman01b}. 

To briefly review, the down-conversion source consisted of a 1.0 mm BBO crystal pumped by roughly 30 mW of the 351.1 nm line of a continuous-wave Argon-Ion laser.  The crystal was cut for degenerate Type-II collinear phase matching and produced pairs of co-propagating, but orthogonally polarized photons at 702.2 nm \cite{rubin94}.
The two photons were separated with an initial polarizing beamsplitter (not shown) and sent along input modes $1$ and $2$ of the $PBS$ shown in Figure \ref{fig:experiment}, where the parity check was performed. In preparation for data collection, the input path lengths and various mode-matching criteria were tested and optimized by studying a variety of standard Shih-Alley \cite{shih86} and Hong-Ou-Mandel \cite{hong87} two-photon interference effects. Typical interference visibilities in this set-up ranged from 75 to 80$\%$.

The horizontal-vertical computational basis was defined by the main $PBS$, and a half-wave retardation plate in input mode $2$ was used to prepare the ancilla photon in the required state 
$|\psi\rangle_{anc} =\frac{1}{\sqrt{2}}(|H_{2}\rangle+|V_{2}\rangle)$ described in Section \ref{sec:theory}.
An additional rotatable half-wave plate in input mode $1$ was used to prepare an arbitrary state of the target photon, 
$|\psi\rangle_{in}\equiv \alpha|H_{1}\rangle+\beta|V_{1}\rangle$.

Note that the polarizing beamsplitter $PBS'$ in the detector package, which was oriented at $45^{o}$ in Figure \ref{fig:qpc}, was implemented in the actual experiment by a conventionally oriented polarizing beamsplitter preceded by a half waveplate used to rotate the reference frame by $45^{o}$. 

As described in the theory of Section \ref{sec:theory}, the post-selection process consisted of accepting the output in mode $1$ only when exactly one photon was found by one of the two detectors $D_{2a}$ or $D_{2b}$.  When this condition was met, the state of the output photon was measured using a polarization analyzer $\theta_{1}$ and an additional single-photon detector $D_{1}$.  Since the probability of having more than two photons in the system at any given time was negligible, the post-selection procedure was equivalent to monitoring the coincidence counting rate between the output ports of the main $PBS$ as a function of $\theta_{1}$. 

A conventional TTL logical OR-gate was used to enable coincidence counting between $D_{1}$ and either $D_{2a}$ or $D_{2b}$, as shown in the upper portion of Figure \ref{fig:experiment}. This OR-gate was part of a custom designed logic board which also enabled the classical information from $D_{2b}$ to control the application of the single-qubit $Z$-correction in output mode $1$. In our experiment, the $Z$-operation (ie. state-dependent $\pi$-phase shift) required by equation (\ref{eq:qpc3}) was implemented on the output mode by a transverse electro-optic modulator (ConOptics Inc. model 360-80/D25 Pockels cell) oriented with its fast and slow axes in the horizontal-vertical basis. The Pockels cell was first DC-biased in such a way that the state of any photons passing through it would remain unchanged. Therefore, if the ancilla photon was detected by $D_{2a}$, the Pockels cell bias voltage was not changed. If, however, the ancilla photon was detected by $D_{2b}$, an accurately amplified TTL signal was used to apply the measured half-wave voltage (roughly 115 V at 702.2 nm) to the unit.  By definition, this half-wave voltage imparts a $\pi$-phase shift on the horizontal polarization component of the state with respect to the vertical component, as required by equation (\ref{eq:qpc3}).

One of the key features of this experiment was a method for storing the output photon while the classical detection signal from $D_{2b}$ was amplified and processed by the Pockels cell driver.  As shown in Figure \ref{fig:experiment} this delay, $\tau$, in output mode $1$ of the quantum parity check was implemented by using a single mode fiber optic delay line (3M Inc. FS-3224). As will be seen in the next subsection, the required time delay of roughly 100 ns was large enough that a free-space delay line would have been impractical in our simple set-up.  The output of the main $PBS$ was launched into and out of the fiber delay line using suitable microscope objectives (fiber couplers) mounted on micro-translation stages, and the coupling efficiency was found to be roughly $50\%$. A standard paddle-wheel polarization controller was used to negate the effects of birefringence induced by the fiber delay line.

\subsection{Required Time Delays}
\label{sec:delays}

In the set-up shown in Figure \ref{fig:experiment}, the total time $\tau_{z}$ required to process the classical signal from $D_{2b}$ and apply the $Z$-operation correction was determined by the operating parameters of the commercially available equipment used in the experiment.  Therefore, the length of the fiber optic cable determining the delay time $\tau$ needed to be carefully chosen.  If $\tau$ was much shorter than $\tau_{z}$ the output photon would pass through the Pockels cell before the half-wave voltage was applied, while if $\tau$ was too long, the half-wave voltage would have been applied and reset before the photon arrived at the Pockels cell.

\begin{figure}
\includegraphics[angle=-90,width=3.25in]{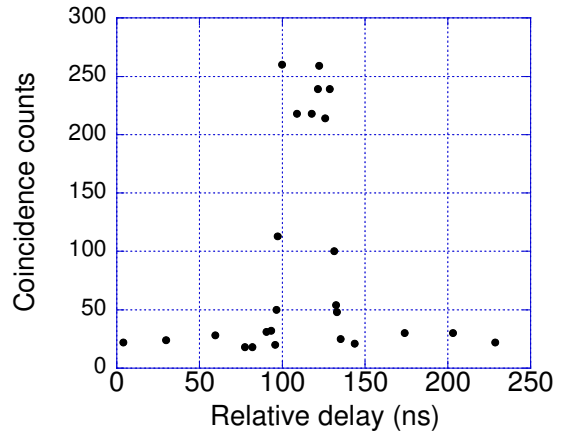}
\vspace*{-.2in}
\caption{Coincidence counting rate between $D_{2b}$ and $D_{1}$ as a function of the relative delay between a fixed fiber optic delay and variable extra electronic delay placed in the Pockels cell driver input channel. For this plot, the waveplates and analyzers were configured in such a way that a coincidence count could only occur if the half-wave voltage was applied to the Pockels cell at the correct time. The results indicate a total system delay time $\tau_{z}$ of roughly 100 ns for our system, which was mostly composed of commercially available components. The data shown also provides an indication of the response of the system to the 33 ns wide output pulses of the single-photon detectors.} 
\label{fig:delays}
\end{figure}

The total system delay $\tau_{z}$ was measured by initially installing a fiber delay line which gave a known optical delay much longer than the expected value of $\tau_{z}$. The state-preparation waveplates and polarizer $\theta_{1}$ were then oriented so that a down-conversion coincidence detection between $D_{1}$ and $D_{2b}$ was only possible if the half-wave voltage was applied to the Pockels cell at the correct time.  Therefore, a plot of the coincidence counting rate as a function of additional electronic delay placed in the Pockels cell driver input channel provided a measure of $\tau_{z}$, as well as an indication of the total system response behavior.  The results of this test are shown in Figure \ref{fig:delays}.  Since the temporal width of the down-converted photon wavepackets and their propagation time through the 10 cm long Pockels cell can be considered negligible, the results of Figure \ref{fig:delays} indicated the need for a minimum delay $\tau$ of roughly 100 ns for our system. This corresponds to a 20 m long fiber optic delay cable, which was subsequently used in the experiments.

The results of tests of the delay times of the individual devices used in our experiment were consistent with total delay indicated in Figure \ref{fig:delays}. The single photon avalanche-photodiode detectors used (Perkin Elmer, SPCM-AQR-12) have built in pre-amplifiers which output a TTL pulse whose width is roughly 33 ns. The time required to produce the leading edge of the output pulse was measured using a triggered high-speed pulsed diode laser and was found to be roughly 18 ns. The overall delay of the Pockels cell, including the driver amplifiers and 3 m triaxial connecting cables, was found to be roughly 38 ns by using a cw light source, crossed polarizers, and a high speed photo-receiver.  The delay induced by our TTL logic board was electronically measured to be 18 ns, and miscellaneous coaxial cables used to connect the devices contributed an extra 26 ns of delay.

For future linear optics quantum computing protocols involving the use of many non-deterministic logic gates in series, one would obviously want to minimize the value of $\tau_{z}$. This can be done by using custom-made high-speed electronics rather than the relatively slow, but commercially available devices used here.  In any event, the results of our tests provide a clear demonstration of the capabilities of a practical feed-forward control system.

\subsection{Results}
\label{sec:results}

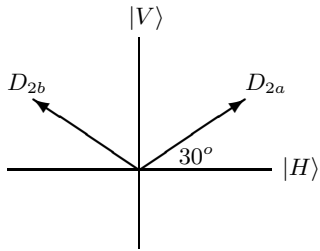
\begin{figure}[b]
\begin{center}
\begin{picture}(110,120)
\put(50,20){\line(0,1){80}}
\put(0,50){\line(1,0){100}}
\put(104,48){$|H\rangle$}
\put(46,105){$|V\rangle$}
\thicklines
\put(50,50){\vector(3,2){40}}
\put(50,50){\vector(-3,2){40}}
\put(90,80){$D_{2a}$}
\put(0,80){$D_{2b}$}
\put(65,52){$30^{o}$}
\thinlines
\end{picture}
\end{center}
\vspace*{-.5in}
\caption{A graphical illustration of the predicted output states of the quantum parity check for an arbitrarily chosen state of the input qubit, 
$|\psi\rangle_{in}= \frac{\sqrt{3}}{2}|H_{1}\rangle + \frac{1}{2}|V_{1}\rangle$, which corresponds to a linear polarization state at $30^{o}$.
The post-selection process described in Section \protect\ref{sec:theory} shows that the input is coherently transferred into the output mode, as desired, if the ancilla photon is detected by $D_{2a}$ .  If, however, the ancilla is registered in $D_{2b}$, a classically controlled $\pi$-phase shift needs to be applied to the horizontal component of the output state. An experimental implementation of this phase shift in real-time is the main result of this paper.}
\label{fig:output} 
\end{figure}

\begin{figure}[t]
\includegraphics[angle=-90,width=3.25in]{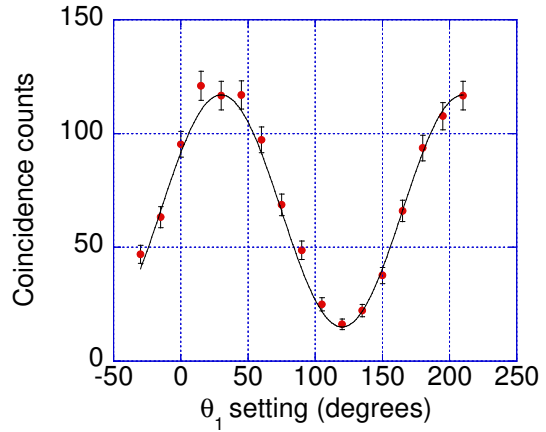}
\vspace*{-.2in}
\caption{Coincidence counts per minute between $D_{1}$ and $D_{2a}$ as a function of the analyzer setting $\theta_{1}$ in the output mode 1. As expected from Figure \protect\ref{fig:output} in this case, the data is consistent with an output state that is linearly polarized at $30^{o}$ and confirms the desired operation of the probabilistic quantum parity check device.  The solid line is plot of a Sine-squared function centered at $30^{o}$ with a visibility defined by the maximum and minimum data values. } 
\label{fig:results1}
\end{figure}

\begin{figure}[b]
\includegraphics[angle=-90,width=3.25in]{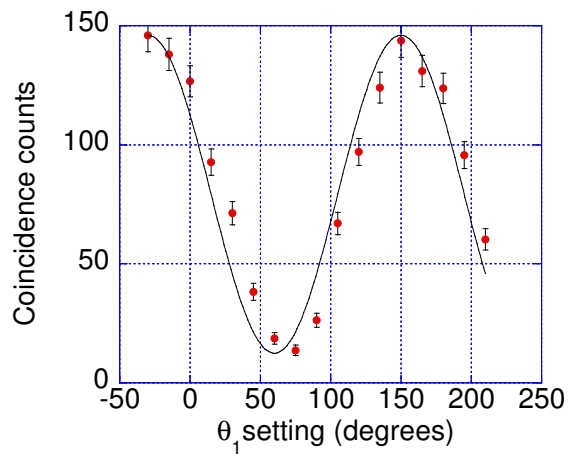}
\vspace*{-.1in}
\caption{Demonstration of the incorrect logical output obtained when the ancilla photon is detected by $D_{2b}$ and the classically controlled $Z$-operation is not applied.  As indicated in Figure \protect\ref{fig:output}, a linearly polarized output at $150^{o}$ is expected in this case.  The slight deviation of the data from the expected value is due to small uncompensated birefringences in our system. In any event, the data clearly shows the need for the classically controlled $\pi$-phase shift on the horizontal component of the output state. } 
\label{fig:results2}
\end{figure}

\begin{figure}[b]
\includegraphics[angle=-90,width=3.25in]{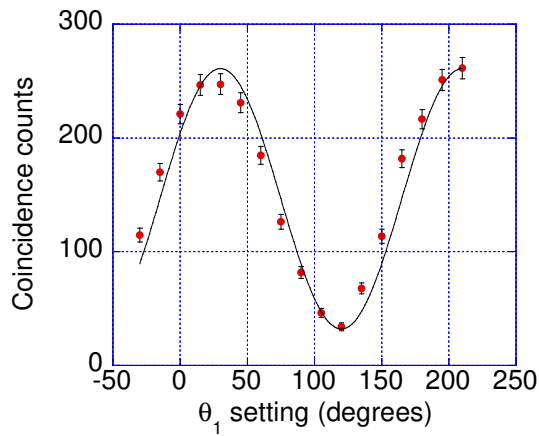}
\vspace*{-.1in}
\caption{Experimental demonstration of feed-forward control in a non-deterministic quantum logic operation.  In this case coincidence counts per minute are recorded between $D_{1}$ and either $D_{2a}$ or $D_{2b}$ as a function of $\theta_{1}$.  The data clearly shows that for those cases in which the ancilla photon is found in $D_{2b}$, the feed-forward system was able to successfully implement the required single-qubit $Z$-operation on the output mode.  As in Figure \protect\ref{fig:results1}, the data shown here is consistent with a linearly polarized output state at $30^{o}$, thereby indicating the successful operation of the quantum parity check.  In addition, the coincidence counting rate here is roughly twice that shown in Figure \protect\ref{fig:results1}, showing that the use of the feed-forward system doubled the success probability of the probabilistic logic operation.} 
\label{fig:results3}
\end{figure}

The results of our demonstration of the use of feed-forward control to increase the success probability of a basic non-deterministic quantum logic operation are summarized in Figures \ref{fig:results1} through \ref{fig:results3}. For these tests of the quantum parity check device, the coefficients $\alpha$ and $\beta$ defining the state of the input qubit were arbitrarily chosen to be:

\begin{equation}
|\psi\rangle_{in} = \frac{\sqrt{3}}{2}|H_{1}\rangle + \frac{1}{2}|V_{1}\rangle
\label{eq:psiin}
\end{equation}

\noindent which corresponds to a linear polarization state of  $30^{o}$.  The output states predicted by equation (\ref{eq:qpc3}) for this choice of the input state are illustrated in Figure \ref{fig:output}. 

The data displayed in Figure \ref{fig:results1} shows the coincidence counting rate between detectors $D_{1}$ and $D_{2a}$ as a function of the setting of the polarization analyzer $\theta_{1}$.  The results clearly show the expected Malus' law dependence on the analyzer setting which is consistent with an output state polarized at $30^{o}$.  Results of this kind were presented in reference \cite{pittman01b} and indicate the non-classical ability of the parity check to coherently transfer the value of the input qubit into the output. 

The data shown in Figure \ref{fig:results2} is analogous to that of Figure \ref{fig:results1}, but displays the coincidence counting rate between detectors $D_{1}$ and $D_{2b}$. For this data run, the TTL input to the Pockels cell driver was intentionally disconnected so that the classically controlled $Z$-correction was not applied. As shown in the illustration of Figure \ref{fig:output}, an uncorrected output state linearly polarized at $150^{o}$ is expected in this case, and the results of Figure \ref{fig:results2} are consistent with this prediction.

The coincidence counting rates at the relevant $\theta_{1}$ settings of $30^{o}$ and $150^{o}$  in Figures \ref{fig:results1} and \ref{fig:results2} average to 131 coincidences per minute (the difference in the maxima of the two plots is due to different overall detection efficiencies in the $D_{2a}$ and $D_{2b}$ channels). By setting the state-preparation waveplates and $\theta_{1}$ to register the maximum coincidence counting rate for otherwise identical experimental conditions, the coincidence rate was found to average 440 coincidences per minute. This number provides an estimate of the total rate of detectable down-conversion pairs entering the system and is in qualitative agreement with the theoretically predicted success probabilities of $\frac{1}{4}$ described in equation (\ref{eq:qpc3}). The discrepancy from the expected value of $\frac{1}{4}$ is due to the effects of birefringence and alignment errors.

Figure \ref{fig:results3} shows the data obtained with the feed-forward control system in full operation, which represents the main result of the paper. The data shows the coincidence counting rate between $D_{1}$ and the output of the TTL OR gate which has $D_{2a}$ and $D_{2b}$ inputs as shown in Figure \ref{fig:experiment}. The data clearly indicates that for those cases in which the ancilla photon was registered by $D_{2b}$, the Pockels cell was able to successfully perform the $Z$-correction in real-time.  As in Figure \ref{fig:results1}, the data shown in Figure \ref{fig:results3} is consistent with an output state polarized at $30^{o}$, indicating the correct operation of the quantum parity check.  In addition to the correct output state, note that the average coincidence counting rate at $\theta_{1}=30^{o}$ is 247 counts per minute.  This data was obtained under the same experimental conditions of that shown in Figure \ref{fig:results1}, and indicates that the success probability of the device was approximately increased from $\frac{1}{4}$ to $\frac{1}{2}$.

\section{Summary and Conclusions}
\label{sec:summary}

In summary, we have experimentally demonstrated the use of feed-forward control to increase the success probability of a basic non-deterministic quantum logic operation.  The experiment involved the use of two polarization-encoded photonic qubits derived from a parametric down-conversion pair.  Classical information describing the detection of one of the photons was fed forward in real-time to a device which then either performed a $Z$-correction or the identity operation to the state of the other photon \cite{chuangneilsen}.  In our system, this feed-forward and correction process was accomplished on a time scale of roughly 100 ns. This particular demonstration involved the use of a probabilistic quantum parity check \cite{pittman01a,pittman01b}, but the techniques presented here are expected to be of general use in many non-deterministic quantum logic operations, as well as a variety of other quantum information processing tasks.

For example, the use of fast feed-forward control is essential in discrete-variable quantum teleportation systems involving parametric down-conversion sources \cite{bouwmeester97,boschi98,kim01,furusawa98}, and significant progress in this  direction has recently been made by DeMartini's group \cite{lombardi01,demartini02}. The techniques presented here may also be useful in, among other things, linear optics based error-correction \cite{bouwmeester01}, entanglement purification \cite{pan01a,yamamoto01,zhao01}, quantum repeaters \cite{kok02}, and quantum relays \cite{jacobs02}.  

Within the context of a complete linear optics quantum computing procedure, the most significant use of feed-forward systems of this type may be for output mode-selection and phase-corrections in a generalized teleportation scheme.  Roughly speaking, one of the basic steps in the original KLM program involves an ingenious generalization of the Gottesman-Chuang protocol \cite{gottesman99} for implementing universal quantum logic gates through teleportation. 
In this procedure \cite{knill01a}, the standard two-photon Bell-state teleportation resource is replaced by an $n$-photon entangled state, $|t_{n}\rangle$, and the Bell-state measurements  \cite{braunstein92} are generalized by a linear optics based Fourier transform operation followed by single-photon detections.
Classical information from the single-photon detectors is fed forward and used to post-select the output mode that contains the correct output state, as well as to apply pre-determined single-qubit corrections such as the $Z$-operation demonstrated here. The probability of an error in the gate operation using teleportation in this way scales as $\frac{1}{n+1}$, which may allow a scalable approach to quantum computing.  A recent ``high-fidelity'' approach  \cite{franson02} reduces the probability of an error to roughly $\frac{2}{n^{2}}$ by using a more optimal entangled ancilla state $|t_{n}\rangle$ and eliminating the use of post-selection (all of the events are accepted). Feed-forward techniques similar to those presented here will play an essential role in either approach. 

The eventual implementation of a linear optics quantum computing procedure will require a system where the number of ancilla is sufficiently large that quantum error correction techniques can be used.  In addition to large numbers of single photons on demand and efficient single photon detectors, it is clear that this procedure will heavily rely on the use of feed-forward control systems of the kind presented here.  Larger scale systems will necessarily involve the use of complicated sequences of classically controlled single-qubit operations and time scales significantly shorter than those described in this basic demonstration. Nonetheless, the techniques and results presented here provide a significant step in that direction.

\begin{center}
{\bf Acknowledgements}
\end{center}
This work was supported by in part by ONR, ARO, NSA, ARDA, and Independent Research and Development funds. We gratefully acknowledge the assistance of M.J. Fitch and useful discussions with M.M. Donegan.



\end{document}